\documentclass[a4paper,11pt]{article}
\usepackage{jinstpub} 
\usepackage{lineno}
\nolinenumbers

\newif\ifshowchanges
\showchangestrue  

\ifshowchanges
    \usepackage[defaultcolor=red]{changes}
\else
    \usepackage[final]{changes}
\fi

\usepackage{subcaption, dcolumn, graphicx}

\usepackage[utf8]{inputenc}
\usepackage[T1]{fontenc}
\usepackage[english]{babel}
\usepackage{hyperref}
\usepackage{textcomp}
\usepackage{xcolor}

\usepackage{amsmath, amssymb}
\usepackage{listings}
\usepackage{adjustbox}
\title{\boldmath A Method for On‑Orbit Calibration of the VLAST‑P Electromagnetic Calorimeter}

\author[a,b]{Jiaxuan Wang}
\author[a,b,1]{Zhen Wang\note{Corresponding author.}}
\author[a,b]{Borong Peng}
\author[a,b]{Renjun Wang}
\author[a,b]{Yunlong Zhang}
\author[a,b]{Zhongtao Shen}
\author[a,b]{Yifeng Wei}
\author[c,d]{Dengyi Chen}
\author[c,d]{Xiang Li}
\author[c,d]{Yiming Hu}
\author[c,d]{and Jianhua Guo}

\affiliation[a]{State Key Laboratory of Particle Detection and Electronics, University of Science and Technology of China,\\
Hefei 230026, China}
\affiliation[b]{Department of Modern Physics, University of Science and Technology of China,\\
Hefei 230026, China}
\affiliation[c]{School of Astronomy and Space Science, University of Science and Technology of China,\\ Hefei 230026, China}
\affiliation[d]{Key Laboratory of Dark Matter and Space Astronomy, Purple Mountain Observatory, Chinese Academy of Sciences, Nanjing 210023, China}

\emailAdd{wangz1996@ustc.edu.cn}

\abstract{
Very Large Area Gamma-ray Space Telescope Pathfinder (VLAST-P), serving as the technology validation satellite for VLAST mission, is designed to observe high-energy solar bursts on orbit.
The CsI electromagnetic calorimeter (ECAL) is one of the key sub-detectors of VLAST-P. In order to study the on-orbit energy calibration method of ECAL, the simulation of VLAST-P was carried out based on Geant4. The results indicated an energy resolution better than 10 \% in the 0.1–5 GeV range, while a dedicated minimum-ionization-particle (MIP) calibration method was developed both to ensure accurate energy reconstruction and to monitor detector stability throughout the on-orbit calibration period.
}


\begin{document}
\maketitle
\flushbottom

\section{Space-Based High-Energy Gamma-Ray Observatories}
\label{sec:global_observatories}
High-energy gamma-ray astronomy has evolved through two main approaches: space-based satellites that directly detect gamma rays and ground-based instruments that observe the particle showers these rays produce in the atmosphere. Ground-based techniques are generally more effective for very high-energy (VHE) gamma rays, as they are less constrained by size and cost, whereas space-based satellites are better suited for observations in the X-ray and high-energy (HE, 30~MeV-30~GeV) ranges~\cite{Gamma-ray_Astronomy}.

The Compton Gamma Ray Observatory (CGRO), one of NASA’s gamma ray observatories in the 1990s, was equipped with four gamma-ray telescopes that collectively covered a broad energy range from the keV to GeV scale. For gamma rays with energies above 10 MeV, where pair production becomes the dominant interaction mechanism, CGRO's high-energy observations were primarily performed by the Energetic Gamma Ray Experiment Telescope (EGRET)~\cite{Thompson2008}. EGRET was designed as a high-energy gamma-ray detector featuring a multilayer spark chamber system for tracking, triggered by a scintillator coincidence system, which was sensitive in the energy range 20 MeV to 30 GeV. Energy measurement was carried out using a total absorption shower counter based on NaI(Tl) crystal, with an overall thickness of 8 radiation length. EGRET achieved an energy resolution of 20–25\% across most of its sensitive energy range~\cite{EGRETCalib}.
 
The Astrorivelatore Gamma ad Immagini LEggero (AGILE) was developed by the Italian Space Agency (ASI) for high-energy astrophysics research. The AGILE payload included four active detectors that together covered a wide energy range, from hard X-rays to gamma rays. Among them, the gamma-ray imaging detector was made up of a segmented anti-coincidence system, a silicon tracker, and a non-imaging CsI(Tl) mini-calorimeter~\cite{Vercellone2024pgh}.

The Large Area Telescope (LAT) onboard the Fermi Gamma-ray Space Telescope is a pair-conversion gamma-ray detector with a wide field of view (2.4 sr) and an energy range from below 20 MeV to more than 300 GeV~\cite{FermiLAT}. It consists of a silicon-strip tracker, a hodoscopic CsI(Tl) calorimeter, and a segmented anti-coincidence detector. The LAT is designed to measure photon direction, energy, and arrival time while rejecting cosmic-ray background. Since its launch, it has enabled all-sky surveys, source localization, spectral measurements, and studies of gamma-ray bursts, diffuse backgrounds, and potential dark matter signals.

In the near future, the Very Large Area gamma-ray Space Telescope (VLAST)~\cite{2022AcASn..63...27F,Pan_2024} is a proposed mission designed to observe gamma rays across a broad energy spectrum, from MeV to TeV. It will utilize two primary detection methods: Compton scattering and electron-positron pair production. The telescope will operate in low Earth orbit, conducting a detailed survey of gamma-ray sources. Its instrumentation includes: an anti-coincidence detector (ACD) to filter out background events, a tracker detector that also functions as a low-energy calorimeter and a high energy imaging calorimeter (HEIC) for precise measurements.

The VLAST detector system builds on previously established gamma-ray detector designs while making important upgrades. First, the ACD blocks unwanted charged particle events and reduces interference from high-energy events. The silicon tracker and low energy gamma-ray detector (STED) uses thin CsI tiles instead of traditional tungsten plates. This change lets the telescope detect both Compton scattering and pair production events while accurately tracking gamma-ray paths. Finally, the HEIC performs three main functions: it measures particle energies, studies particle showers in detail, and helps distinguish electron events from proton events while estimating their directions.

\section{Very Large Area Gamma-ray Space Telescope Pathfinder}
\label{sec:vlastp_overview}

The VLAST-Pathfinder (VLAST-P), a compact version of the VLAST detector, is scheduled for launch in 2026. As a pathfinder mission, it serves dual purposes: validating key technologies for the full VLAST mission, and conducting groundbreaking solar observations. The primary scientific objectives of the VLAST-Pathfinder mission are to address critical gaps in solar high-energy observations by detecting gamma rays in the 100~MeV-5~GeV energy range, capturing high-energy proton signatures during solar flares, and conducting coordinated observations with China's ASO-S ("Kuafu-1") solar satellite to enable comprehensive multi-wavelength studies. A key focus will be on investigating the fundamental relationship between electron and proton acceleration mechanisms during flare events, which remains poorly understood in current solar physics research.

\begin{figure}[htb]
\centering
\includegraphics
  [width=0.6\hsize]
  {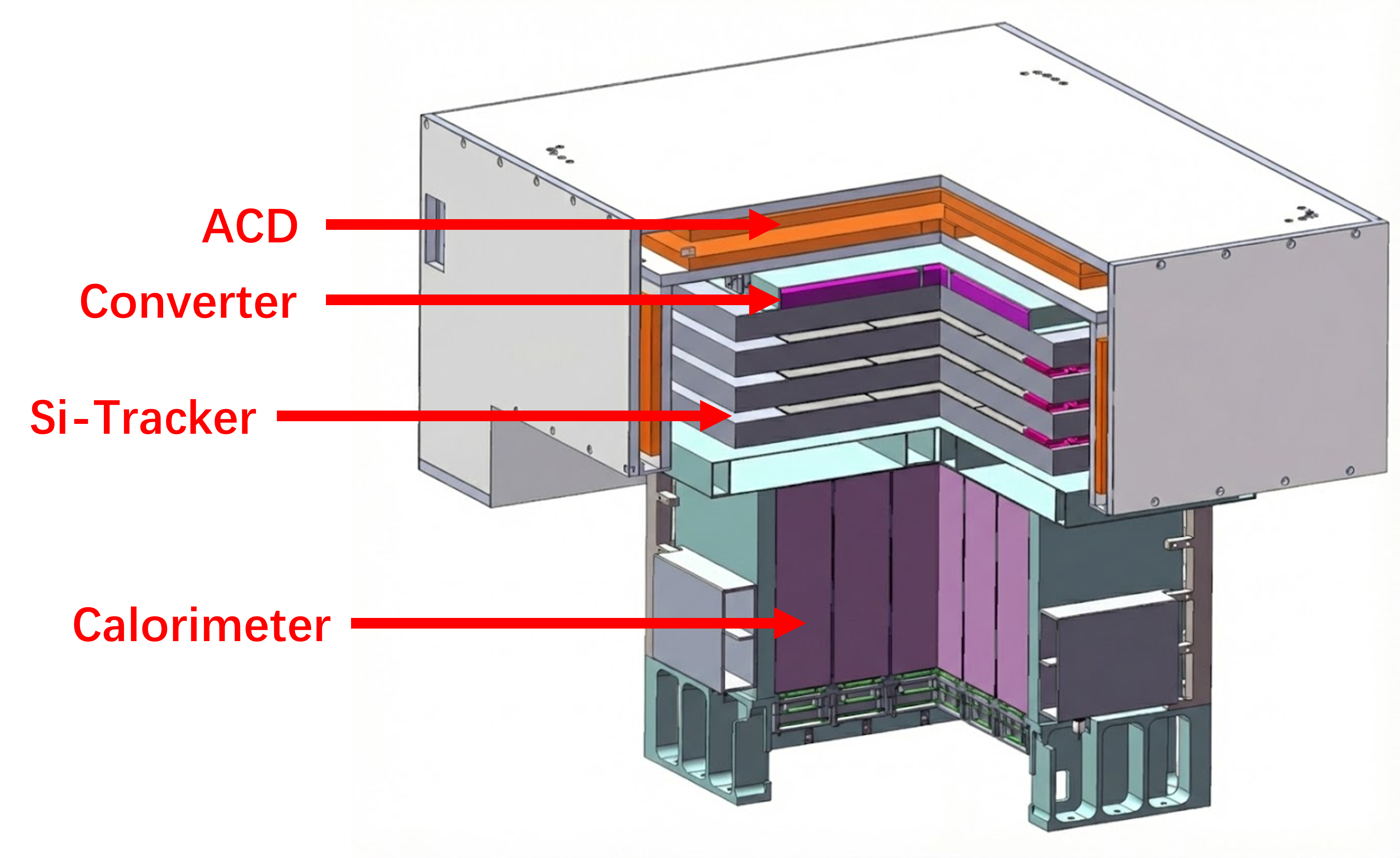}
\caption{The schematics of the VLAST-P detector system. It inherits the baseline design of the VLAST detector system. ACD serves to distinguish charged particles from gamma rays. The tracker is used to measure the tracks of charged cosmic rays or converted electron pairs. The ECAL is used to measure the energy of particles entering the crystals.}
\label{fig:VLASTP}
\end{figure}

Figure~\ref{fig:VLASTP} shows the schematics of the whole detector system. It maintains the design of VLAST detector system. Its detector system comprises three main sub-detector systems: ACD, tracker and electro-magnetic calorimeter (ECAL). \replaced{The ACD consists of two layers of plastic scintillators: the upper layer is a single plate ($400\,\mathrm{mm} \times 400\,\mathrm{mm}$), while the lower layer is segmented into four smaller tiles ($200\,\mathrm{mm} \times 200\,\mathrm{mm}$) arranged in an overlapping configuration to minimize dead zones. The signals are read out by PMTs coupled with wavelength-shifting fibers. This design is driven by two primary technical objectives: to achieve effective separation between charged particles and gamma rays (with a charged particle detection efficiency exceeding 99.9\%), maintain optimal performance at high event rates (up to $10\,\mathrm{kHz/m^2}$).}{} \replaced{The VLAST-P tracker consists of two main components: a $15\,\mathrm{mm}$ thick CsI converter ($\sim 0.8\,X_0$) and a silicon tracking section. It serves a dual purpose: acting as an active medium for photon conversion into electron-positron pairs, and reconstructing the incident direction by measuring the pair trajectories with high precision. Unlike traditional passive tungsten converters, the active CsI layer measures the energy deposited during conversion. This capability allows for complete energy reconstruction, which is critically important for low-energy particles.}{}\replaced{Following the converter, the silicon tracker consists of six silicon micro-strip layers arranged in an alternating x-y configuration. Each pair of orthogonal strips provides 2D coordinate measurements. By adopting the silicon strip detector technology successfully verified in the DAMPE mission~\cite{AZZARELLO2016378}, this design is able to characterize the electron-positron pairs from photon conversion with a spatial resolution better than $50\,\mu\mathrm{m}$.}{}

\replaced{The main sub-detector for particle energy measurement is the electromagnetic calorimeter (ECAL). Given the stringent constraints on satellite weight and dimensions, the ECAL is optimized to balance performance with available resources. It is composed of 25 CsI(Tl) crystal bars arranged in a $5\times5$ array, as shown in the lower part of figure~\ref{fig:VLASTP}. Thallium-doped cesium iodide (CsI(Tl)) is selected as the sensitive material due to its high light yield and excellent uniformity. Each crystal measures $60\,\mathrm{mm}\times60\,\mathrm{mm}\times200\,\mathrm{mm}$. With a radiation length ($X_0$) of $1.86\,\mathrm{cm}$ for CsI(Tl), the longitudinal depth of approximately $10.8\,X_0$ is sufficient to contain a significant fraction (about 93\% for 1 GeV photons) of the electromagnetic shower. Transversely, with a Molière radius of $3.57\,\mathrm{cm}$, the $300\,\mathrm{mm} \times 300\,\mathrm{mm}$ cross-section ensures effective containment of the shower core.}{The main sub-detector for particle energy measurement is the electromagnetic calorimeter (ECAL). The detector is optimized to cover an energy range of $100\,\mathrm{MeV}$ to $5\,\mathrm{GeV}$, although the ECAL design is limited by the area and weight available on the satellite. It is built with 25 CsI(Tl) crystal bars arranged in a $5\times5$ array, as shown in the lower part of figure~\ref{fig:VLASTP}. Thallium-doped cesium iodide is chosen as the sensitive material because of its high light yield and good uniformity. The size of each CsI(Tl) crystal is $60\,\mathrm{mm}\times60\,\mathrm{mm}\times200\,\mathrm{mm}$.
With a radiation length ($X_0$) of $1.86\,\mathrm{cm}$ for CsI(Tl), the longitudinal depth of the ECAL corresponds to approximately $10.8\,X_0$, which is sufficient to contain a large portion of the electromagnetic showers from GeV photons.} The crystals are wrapped with $\mathrm{TiO_2}$ reflective films to fully collect the scintillation photons. The crystals are separated by a mechanical gap of $1\,\mathrm{mm}$; thus, the total active volume is approximately $300\,\mathrm{mm}\times300\,\mathrm{mm}\times200\,\mathrm{mm}$.

\replaced{The calorimeter is designed to measure incident photons with energies up to 5\,GeV. For a 5\,GeV photon, simulations indicate that the total energy deposited in the full ECAL approaches 5\,GeV. The maximum energy deposition observed in a single crystal is approximately 4.1\,GeV, which determines the required dynamic range of the front-end readout electronics, since each channel must accommodate the largest possible energy deposited in an individual crystal.}{The required readout energy range is designed to cover the energy range up to $5\,\mathrm{GeV}$. To measure incident photons up to $5\,\mathrm{GeV}$, simulations indicate a maximum energy deposition of approximately $4.1\,\mathrm{GeV}$ in a single crystal, which determines the upper limit of the measurement range.} At the low-energy end, the energy threshold of single crystal in simulation is set to $3\,\mathrm{MeV}$; therefore, the electronic noise must be controlled to meet the threshold. As a result, the readout system is required to cover a wide dynamic range \replaced{from $3\,\mathrm{MeV}$ up to approximately 4.1\,GeV}{from the $3\,\mathrm{MeV}$ up to the $4.1\,\mathrm{GeV}$ saturation level.}

\replaced{}{The main role of the detector for particle energy measurement is the electromagnetic calorimeter (ECAL). The ECAL design is limited by the area and weight available on the satellite. It is built with 25 CsI crystal bars placed evenly in a $5\times5$ array, as shown in the lower part of figure~\ref{fig:VLASTP}. Thallium-doped cesium iodide is chosen as the sensitive material because of its high light yield and good uniformity. The size of each CsI(Tl) crystal is $60~mm\times60~mm\times200~mm$;therefore, the total size of the VLAST-P ECAL is approximately $300~mm\times300~mm\times200~mm$.The depth and area are optimized to fully exploit the detection capability given weight and space constraints.}

\section{Simulation and Performance Evaluation of the VLAST-P Detector}
\label{sec:simulation}
\subsection{Geant4 Simulation Framework}
In this work, the Geant4 toolkit~\cite{ALLISON2016186,AGOSTINELLI2003250} is employed to perform detailed Monte Carlo simulations of particle interactions within the VLAST-P payload. 
Geant4 is a comprehensive framework for modeling the transport and interaction of high-energy particles with detector materials, incorporating both electromagnetic and hadronic processes across a broad energy range. 
Its modular architecture allows for the accurate implementation of geometry, material composition, and detector response, ensuring a realistic description of the experimental setup. 
For the present study, the simulation focuses on the response of the CsI(Tl)-based electromagnetic calorimeter to cosmic-ray protons, helium nuclei, and solar $\gamma$-rays, which constitute the dominant sources of background and calibration signals in orbit. 
The use of Geant4 thus enables quantitative evaluation of detector performance, including energy deposition, shower development, and calibration stability, providing critical input for both event reconstruction and on-orbit calibration strategies.\\

\subsection[Detector Simulation Setup]{Detector Simulation Setup}
\begin{figure}[htb]
\centering
\includegraphics
  [width=0.95\hsize]
  {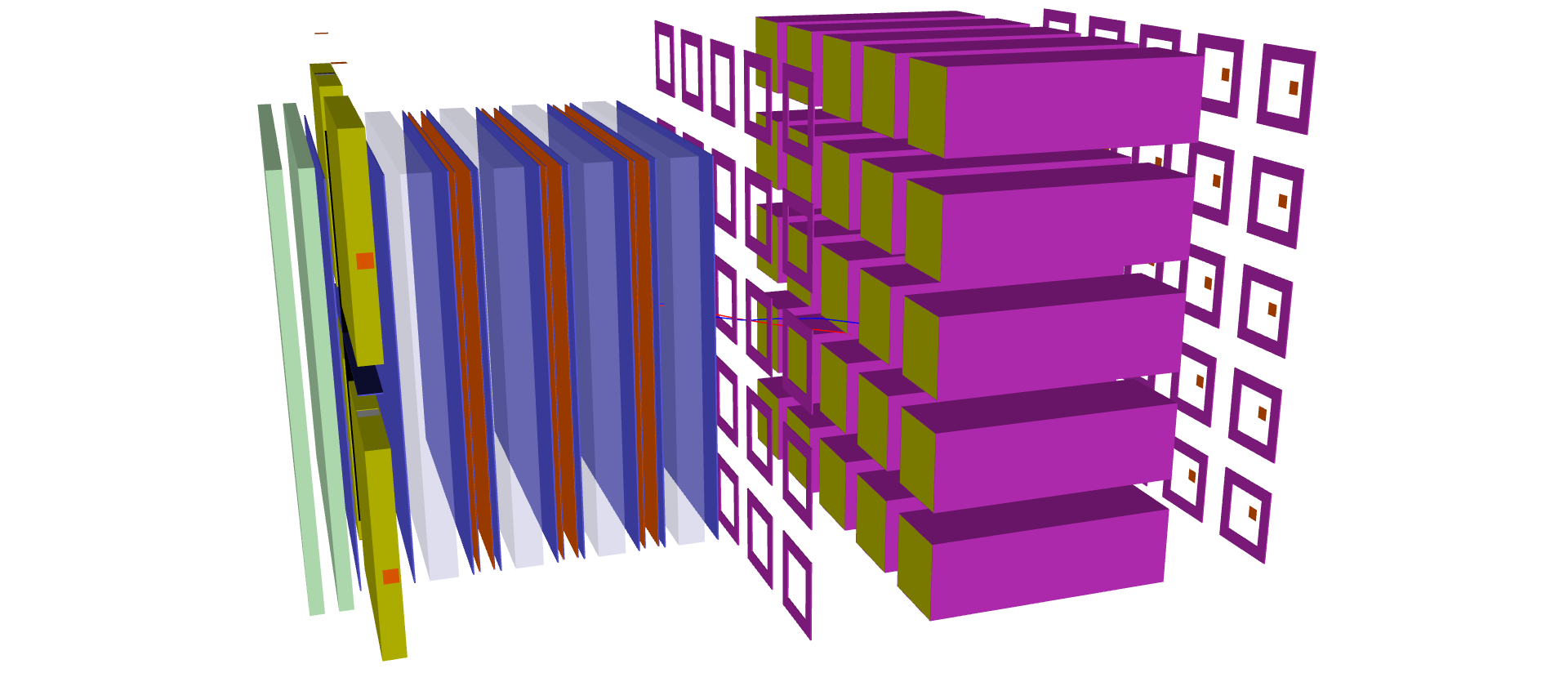}
\caption{Exploded 3D visualization of the VLAST-P detector system, generated using Geant4 and JSROOT.}
\label{fig:VLASTP_MC}
\end{figure}

Figure~\ref{fig:VLASTP_MC} shows the exploded view of our simulated detector system drawn with front-end tool JSROOT~\cite{JSROOT}. The components are displayed in separate layers to illustrate the internal structure. As shown in the picture from left to right, the two outermost plates (shown in light green) are the ACD layers, which are positioned at the front to effectively distinguish charged particles from photons. Next to the ACD, four CsI(Tl) plates (shown in yellow) are placed before the six silicon tracker layers (depicted in blue and red layers). Specifically, four CsI(Tl) plates serve as the converter while the six silicon microstrip plates (used to measure x and y coordinates) provide position information with 3 oriented along the x-axis and 3 along the y-axis. Each of the two adjacent plates constitutes a layer that provides position information for the hit. Finally, on the far right, the ECAL is visualized as the $5\times5$ array of purple crystal bars, which are used to measure the energy of particles entering the crystals. Detailed mechanical and electronic components are also visible: the small red squares attached to the crystal ends are APDs, while the purple frame-like structures correspond to the silicone rubber pads designed for shock absorption. Inclusion of these fine details ensures a more realistic representation of the detector geometry.

As mentioned before, CsI(Tl) crystals are used in the structure of ECAL. It comprises a $5\times5$ array of crystal arrays. Each crystal is wrapped with a titanium dioxide ($\mathrm{TiO_2}$) reflective coating, which serves two purposes: preventing hygroscopic degradation of the crystal and reflecting internally generated scintillation light toward the avalanche photodiode (APD) on one end for optimal photon collection efficiency. The simulation framework enables detailed modeling of detector materials in the VLAST-P detector system, including its carbon fiber support structure and embedded silicone rubber buffer pads. The carbon fiber frame ensures structural stability during launch and orbital operations, while the silicone rubber layers mitigate micro-vibrations that could probably affect crystal performance. 

In addition to the geometry and materials, the simulation incorporates a digitization \replaced{process}{step} to process the raw energy deposition to obtain a more realistic reconstructed energy value. The process begins by converting the deposited energy into the number of photoelectrons based on the light yield of the CsI(Tl) crystals. Subsequently, Gaussian smearing is applied to the signal to account for both the statistical fluctuations and the non-uniformity of the crystal response, where the latter was determined to be $2\%$ based on our single-crystal cosmic ray tests. The total signal also includes the ionization energy deposited by particles passing directly through the APDs. Finally, an electronic noise level of approximately $1\,\mathrm{MeV}$ is superimposed on the signal to mimic the realistic detector performance.

\subsection{Performance and Validation}
\begin{figure}[htb]
    \centering
    \includegraphics[width=0.95\hsize, height=6.5cm, keepaspectratio]{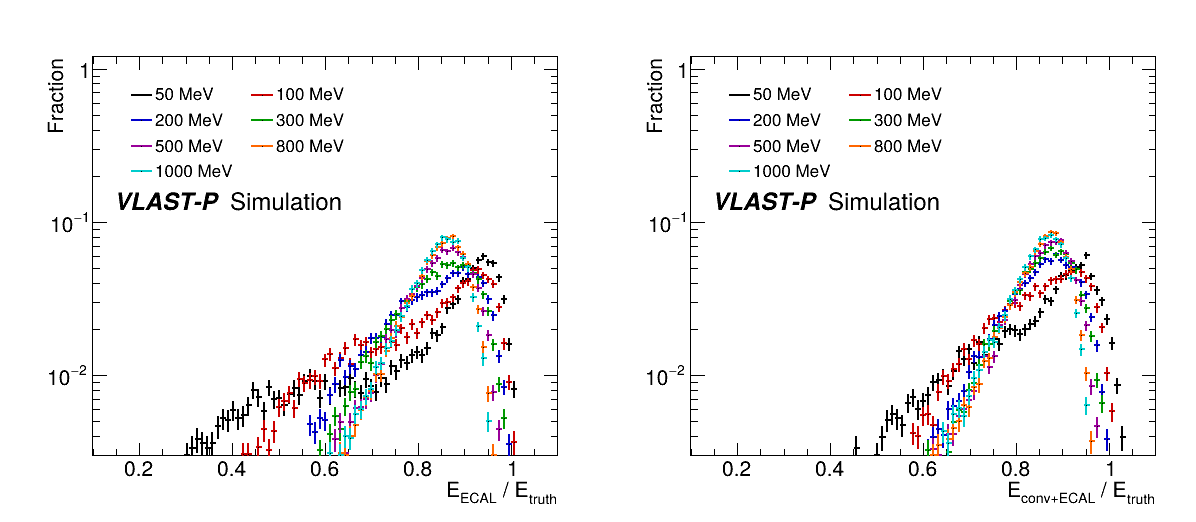}
    \caption{Comparison of reconstructed energy using only the ECAL (left) versus including the active converter's deposit (right).}
    \label{fig:ECALvsConvECAL}
\end{figure}

\begin{figure}[htbp]
    \centering
    \begin{subfigure}{0.32\textwidth}
        \centering
        \includegraphics[width=\linewidth]{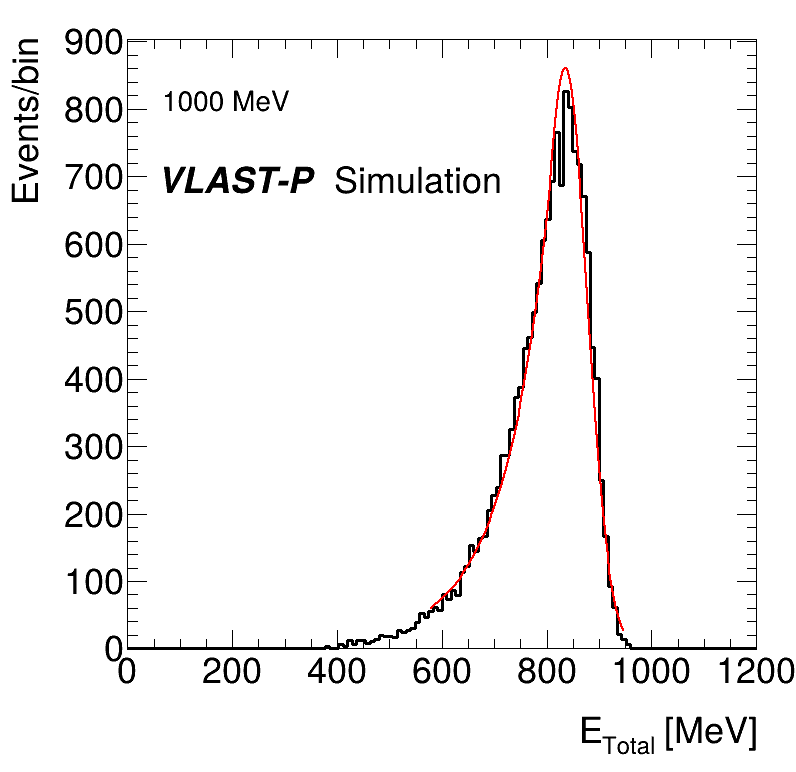}
        \caption{Example fit at 1000 MeV}
        \label{fig:ECAL_c}  
    \end{subfigure}
    \hfill
    \begin{subfigure}{0.32\textwidth}
        \centering
        \includegraphics[width=\linewidth]{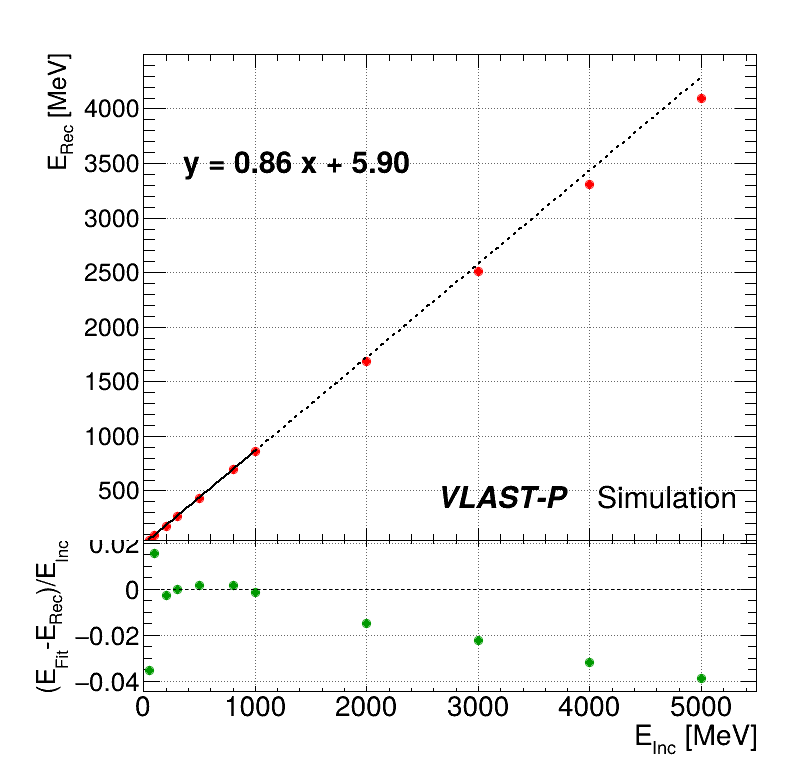}
        \caption{Energy linearity}
        \label{fig:ECAL_a}  
    \end{subfigure}
    \hfill
    \begin{subfigure}{0.32\textwidth}
        \centering
        \includegraphics[width=\linewidth]{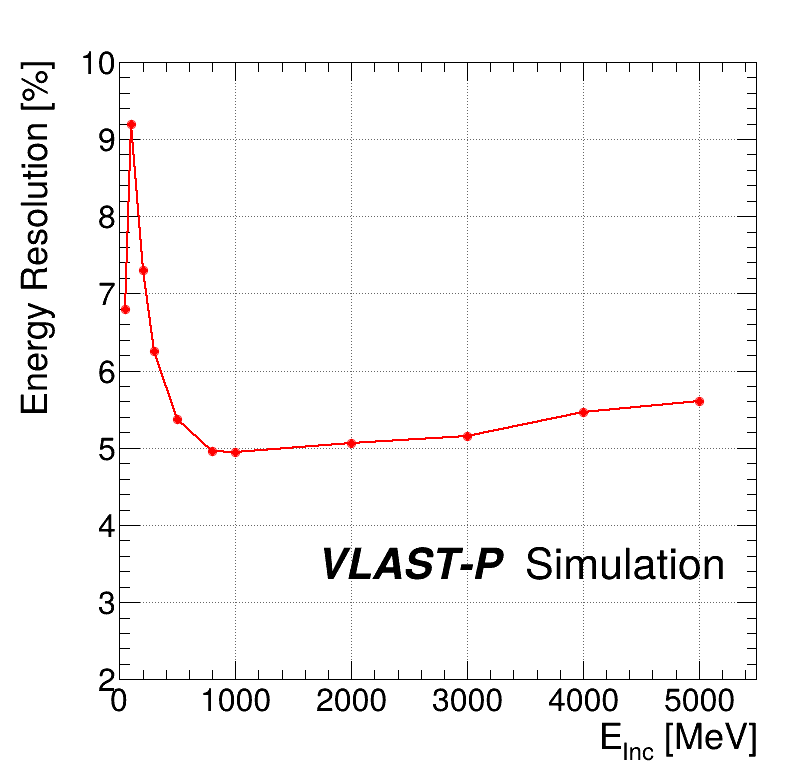}
        \caption{Energy resolution}
        \label{fig:ECAL_b}  
    \end{subfigure}
    \caption{Performance of the VLAST-P ECAL: estimated energy linearity and energy resolution.\replaced{ Only events with photon conversion process are selected for analysis.}{}}
    \label{fig:ECAL_perf} 
\end{figure}

By employing a full Geant4-based simulation, the intrinsic energy response of the calorimeter to monoenergetic photons was evaluated. Gamma rays with energies ranging from 50~MeV to 5~GeV were simulated at normal incidence to the detector surface. The primary particles were emitted from a planar source of size $150~\mathrm{mm} \times 150~\mathrm{mm}$, positioned 10~mm away from the center of the ACD's front surface. The particles were incident perpendicularly to the detector surface. A total of $1 \times 10^5$ events were generated to ensure sufficient statistical precision for each incident energy point. The energy of each event is reconstructed \replaced{by summing the energy deposits in the crystal with the maximum signal and its neighboring crystals, together with the energy deposited in the active converter. As shown in
figure~\ref{fig:ECALvsConvECAL}, this combined energy reconstruction is important because a non-negligible fraction of the incident energy, particularly for low-energy gamma rays, is deposited in the converter before the electromagnetic shower develops in the ECAL. If only ECAL signals are used, the reconstructed energy is systematically underestimated also leading to a visible low-energy tail at 50 and 100 MeV. By including the energy deposited in the converter, this bias is largely corrected, and the reconstructed energy distribution shifts closer to unity, improving the accuracy of the low-energy measurement. }{by summing the energies of adjacent crystals surrounding the one with the maximum deposition.} \replaced{}{The VLAST-P tracker consists of two components: an active converter layer, where incident gamma rays are expected to undergo pair production, and a silicon tracking section that records the subsequent electron-positron trajectories with high spatial resolution. The tracker serves two essential functions: (i) photon conversion, by providing an active medium in which gamma rays can convert into an electron--positron pair, and (ii) direction reconstruction, by accurately measuring the pair trajectories to
infer the incoming photon’s direction.} In the context of the ECAL simulation, tracker information is used to identify whether the photon has converted in the tracker; \replaced{to evaluate the energy reconstruction performance for the pair-production channel, we specifically select events where the photon converts into an electron-positron pair within the CsI converter. The total reconstructed energy is defined as the sum of the energy deposited in the active CsI converter and the energies measured by ECAL crystals ($E_{\mathrm{rec}} = E_{\mathrm{CsI}} + E_{\mathrm{ECAL}}$). This ensures that the contribution of the active converter to the total energy measurement is properly included in the evaluation.}{events that convert upstream are either rejected or treated separately, ensuring that the energy response to calorimeter is evaluated only for photons entering the ECAL as intended.}
As illustrated in figure~\ref{fig:ECAL_c}, the energy distribution at each incident energy point is fitted with a Crystal Ball function~\cite{PhysRevD.34.711} to extract the mean reconstructed energy and the full width at half maximum (FWHM). The energy resolution \( R \) is calculated from the FWHM of the energy peak as follows:
\begin{equation*}
R = \frac{\mathrm{FWHM}}{2.35 \times \mu} \, \text{\cite{knoll2010radiation}}
\end{equation*}

where \( \mu \) is the fitted energy of the peak. The factor 2.35 converts the FWHM to the standard deviation, assuming a Gaussian distribution.

The performance of the VLAST-P ECAL was evaluated in terms of both energy linearity and resolution for incident gamma rays. As shown in figure~\ref{fig:ECAL_a}, the measured energy deposited in the CsI(Tl) scintillator exhibits a nearly proportional response to the true incident energy over the range from $50\,\mathrm{MeV}$ to $5\,\mathrm{GeV}$, demonstrating good energy linearity with less than \replaced{5}{10}\% deviations. \replaced{The resolution curve, illustrated in figure~\ref{fig:ECAL_b}, shows a sharp decrease in the low-energy range, \replaced{dropping from approximately $10\%$ to about $5\%$ at $1000\,\mathrm{MeV}$.}{dropping from over $20\%$ to roughly $6\%$ at $1000\,\mathrm{MeV}$.} Above $1\,\mathrm{GeV}$, the improvement slows down, and the resolution stabilizes at a plateau of approximately $6\%$ up to $5\,\mathrm{GeV}$, indicating stable performance across the high-energy range.}{Correspondingly, the energy resolution, illustrated in figure~\ref{fig:ECAL_b}, remains around 5\% across a wide energy range for gamma rays normally incident on the detector. This is due to the high stopping power of the CsI(Tl) crystals, which absorb most of the incident energy.} Consequently, only a small fraction of energy leakage occurs even at higher energies, ensuring accurate energy reconstruction.

\replaced{}{The resolution curve shows a slight dip around 500 MeV, which corresponds to an optimal balance between sufficient energy deposition and minimal fluctuation in shower containment. At this point, the electromagnetic shower is largely contained within the detector volume, and stochastic fluctuations are relatively suppressed.} \replaced{The longitudinal containment of the ECAL is defined by its thickness of 10.8\,$X_0$. 
Monte Carlo simulations with 100\,MeV and 1\,GeV gamma rays indicate that approximately 93\% of the incident energy is contained within this 20-cm-long CsI crystal.
The remaining leakage is attributed to the finite detector depth, and future reconstruction improvements will include a dedicated leakage correction for high-energy events. The transverse containment is governed by the Molière radius of CsI, which is 3.57\,cm, and it is generally accepted that 95\% of the shower energy is contained within two Molière radii. When the shower maximum is located within the central $3\times3$ crystal array, the ECAL geometry is sufficient to contain nearly all transverse shower energy deposition.} {As the incident energy increases beyond this point, the resolution gradually worsens due to increasing longitudinal and transverse shower leakage, which leads to larger fluctuations in the reconstructed energy and thus a degradation in resolution.}\\

\section{On-Orbit Calibration of the Electromagnetic Calorimeter}
\label{sec:ecal_calibration}
\subsection{Simulation Methodology}
\label{sec:simulation_methodology}
Primary cosmic rays consist of about 90\% protons, 10\% helium nuclei, and other heavy ions.This is in contrast to the cosmic ray flux at sea level, which is dominated by secondary muons Thus, calibration of the sensitive units in space-based detectors is most commonly performed using minimum ionizing protons from primary cosmic rays. This simulation aims to investigate the MIP-induced signal response in a 5 × 5 crystal array detector with GEANT4, through the development of a backtracing database grounded in a realistic geomagnetic field model.

Backtracing is generally easier and more efficient than forward tracing, as it launches from the observation point and avoids sampling the vast incident phase space~\cite{osti_4784046}. A comprehensive global geomagnetic backtracing database, named GeoMagFilter, has been constructed in Ref.~\cite{CHEN20255450}. It was generated using specialized software that simulates particle trajectories via an eighth-order Runge-Kutta numerical integration method within the International Geomagnetic Reference Field (IGRF-13) model. The database is constructed with three parameters: particle rigidity, zenith angle, and azimuth angle. Particle rigidity is defined as momentum divided by the charge quantity of the particle to represent how difficult it is for a particle to be deflected by a magnetic field. The latter two parameters can be calculated based on geographic longitude and latitude, which are typically accessible as basic orbital information of the satellite. The sampling grid spans longitudes from $0^\circ$ to $360^\circ$ in $10^\circ$ intervals and latitudes from $-75^\circ$ to $+75^\circ$ in $5^\circ$ steps, covering all possible orbital positions of both the International Space Station (ISS) and the China Space Station (CSS). Directional angles are sampled uniformly on a spiral in $4\pi$ steradians with an angular spacing of 0.03 radians, resulting in approximately 13,963 distinct directions. A binary value is assigned to indicate whether a particle can successfully trace back beyond the geomagnetic field (1 for successful backtracing, 0 for failure, e.g., when the trajectory terminates within $10^5$ km).

Given the orbital altitude of 500~km consistent with satellite, the database indicates the rigidity a particle can be backtraced successfully when the satellite is at a certain position. By dividing the $4\pi$ steradians, distribution of proton flux of each distinct direction can be obtained. Using a dedicated full-detector simulation, the incident direction and rigidity of particles are sampled to mimic a cosmic ray source. In order to generate events efficiently, the following method is adopted in the simulation.

First, a random center point is selected within the ECAL region and treated as the center of a spherical cosmic ray source. A direction is then randomly chosen from the backtracing database to define the trajectory of incoming particles. By setting the radius of the cosmic ray source to 1 meter, the simulated sphere can cover the entire detector region. However, in a realistic detector system, an event is only triggered when all sub-detectors register signals. Most of the events generated using the above method may still fail to meet the trigger conditions. To improve the efficiency of event simulation, a filtering process is further applied. Before generating a particle, its truth trajectory is calculated. If the extension of the initial trajectory does not intersect all sub-detectors, the event simulation is terminated. With this technique applied, realistic and detector-specific cosmic ray event simulations can be efficiently produced.

\subsection{Event Selection}

\begin{figure}[htb]
    \centering
    \includegraphics[width=0.95\hsize, height=6.5cm, keepaspectratio]{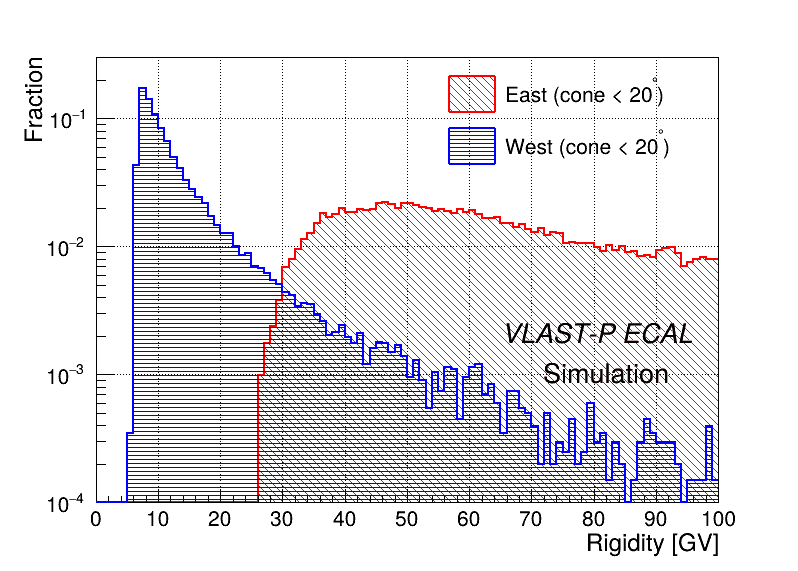}
    \caption{Particle rigidity spectrum from a 20 degrees cone along east direction and west direction}
    \label{fig:East_West}
\end{figure}

\replaced{Calibration events are selected using a strict four-fold coincidence logic requiring valid hits in both ACD layers, the converter, and the ECAL. The specific energy thresholds are set to $0.6\,\mathrm{MeV}$ for each ACD layer, $2.52\,\mathrm{MeV}$ for single converter tile, and $36\,\mathrm{MeV}$ for single ECAL crystal, ensuring the effective rejection of background noise.}{Due to arbitrary incident direction of cosmic ray particles, it is necessary to distinguish the events that can be used for calibration. First, according to the trigger logic, during the MIP calibration process, it is required that the hit signals from multiple detectors be combined using a logical AND operation. This ensures that only events where several detectors are simultaneously triggered are selected for calibration. A more specific set of thresholds is applied in the simulation, requiring a logical AND signal between the two ACD layers, as well as valid energy depositions in both the converter region of the tracker and the calorimeter. An event is kept if at least two units in two ACD layers, one for each, have more than 0.6 MeV of deposited energy, one converter unit has more than 2.52 MeV, and one ECAL cell has more than 36 MeV. These thresholds help to ensure that the signals come from real particle interactions rather than random background.}
The input spectrum used to simulate the realistic spatial cosmic ray environment is generated from the database described earlier. From these input spectra,\replaced{ a pronounced east–west effect is observed for positive particles}{ a pronounced east–west effect is observed}, as shown in figure~\ref{fig:East_West}.

This effect arises due to the influence of the geomagnetic field on charged particle trajectories and is consistent with physical expectations for low-energy cosmic rays. To ensure that the ECAL records high-purity signals originating from genuine particle interactions, a stringent event selection procedure is applied. 

For particles penetrating the detector at large incident angles, the path length calculation can be complicated by shower development. To determine the optimal selection criterion, we analyzed the hit cell multiplicity distributions for both showering and non-showering events. Based on this study, a threshold of fewer than 6 hit cells was established. This cut effectively discriminates against showering events, ensuring that the selected tracks have well-defined path lengths within the crystals.
\replaced{Furthermore, a correction for the path length has been applied in this step by evaluating the true trajectory using the incident angle and initial position of the particle, as measured by the tracker located upstream of the ECAL.}{Furthermore, a correction for the path length has been applied in this step by evaluating the true trajectory using the incident angle and initial position of the particle. \replaced{A tracker is located upstream of the ECAL.}{There lies a tracker in front of the ECAL.}} In the real on-orbit calibration, with the reconstruction algorithm of the tracker, the incident angle can also be used to select events. Additionally, in some events, it is inevitable that some hits come from a very short path length. To exclude these, hits with a path length less than 2 cm are not counted.

\subsection{Performance}
The selection criteria are primarily designed to reject events caused by particles entering at large incident angles, thereby retaining events suitablefor calibration purposes. \replaced{As presented in Table~\ref{tab:cutflow}, each selection stage leads to a substantial reduction in the number of surviving events. The table lists both the selection efficiency at each step and the cumulative efficiency. Starting from the full simulated dataset, the event count decreases sequentially as increasingly restrictive criteria are applied. Logic trigger selects events satisfying the detector’s trigger conditions. ECAL hit number is applied to suppress cluster shower events. Incident direction requires that the particle trajectory reconstructed by the tracker in the simulation is within 20° of the z-axis, ensuring nearly normal incidence. After the complete set of source-filter selections described in Section~\ref{sec:simulation_methodology}, only 2.95\% of the total simulated events remain.}{As presented in table~\ref{tab:cutflow}, each selection stage leads to a substantial reduction in the number of surviving events. The table also lists the selection efficiency at each stage and the cumulative efficiency up to that point. Starting from the full simulated dataset, the event count decreases step-by-step, ultimately retaining only 2.95\% of the total events with a source filter mentioned in Section~\ref{sec:simulation_methodology}.} This result demonstrates that the selection strategy is effective in suppressing showering events while preserving a clean sample of physical events suitable for further analysis.

\begin{table}[htbp]
    \centering
    \caption{Primary proton cut flow table}
    \begin{tabular}{c|c|c|c}
        \hline
        $\mathrm{Cut}$ & $\mathrm{Event}_{\mathrm{selected}}$ & $\mathrm{Eff}_{\mathrm{Step}}$ & $\mathrm{Eff}_{\mathrm{total}}$ \\
        \hline
        No cut  & 10000000 & 100\% & 100\% \\
        Logic Trigger  & 3819981 & 38.1\% & 38.1\% \\
        ECAL Hit No & 868582 &  22.7\% & 8.68\% \\
        Incident Direction & 295410 & 34.0\% & 2.95\% \\
        \hline
    \end{tabular}
    \label{tab:cutflow}
\end{table}

\begin{figure}[htb]
\centering
\includegraphics
  [width=0.95\hsize, height=6.5cm, keepaspectratio]
  {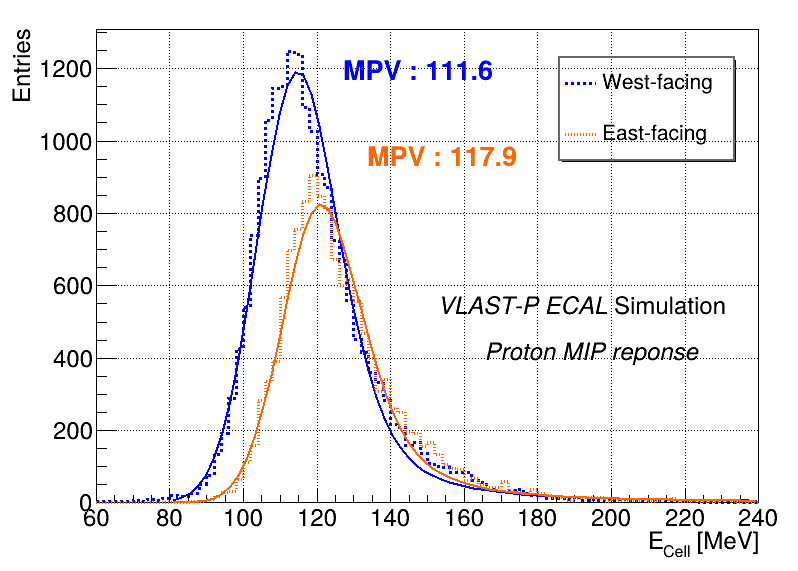}
\caption{Proton MIP response of the VLAST-P ECAL for east- and west-facing configurations}
\label{fig:Cell_East_west}
\end{figure}

On orbit, the satellite can operate in a sun-pointing mode, in which the front side of the detector continuously faces the Sun. Under this configuration, the detector alternately points toward the west and east directions with an approximately equal probability. Since the incident particle rigidity spectrum can differ significantly between the two orientations, the rigidity spectra shown in figure~\ref{fig:East_West} are separately adopted as input distributions in the simulation to evaluate their impact. The analysis indicates that the MPV of the energy deposition for protons arriving from the East and West directions exhibits a \replaced{shift}{discrepancy} of approximately 6\%, as illustrated in figure~\ref{fig:Cell_East_west}.

A path length correction is applied in the analysis to compensate for the effect of particle incident angles on the measured energy deposition. 
Particles entering the detector at oblique angles traverse a longer path through the scintillator material and may penetrate multiple crystal bars, resulting in larger apparent energy depositions compared with perpendicular incidence. 
\replaced{}{Without this correction, the reconstructed energy for large-angle events would be systematically underestimated due to the significant number of oblique incidences.}
After applying the path length correction, the energy spectrum exhibits a Landau-like shape, and the MIP peak position becomes more consistent across different angular ranges, improving both the accuracy and stability of the calibration.

\begin{figure}[htb]
    \centering
    \begin{subfigure}{0.48\textwidth}
        \centering
        \includegraphics[width=\linewidth, keepaspectratio]{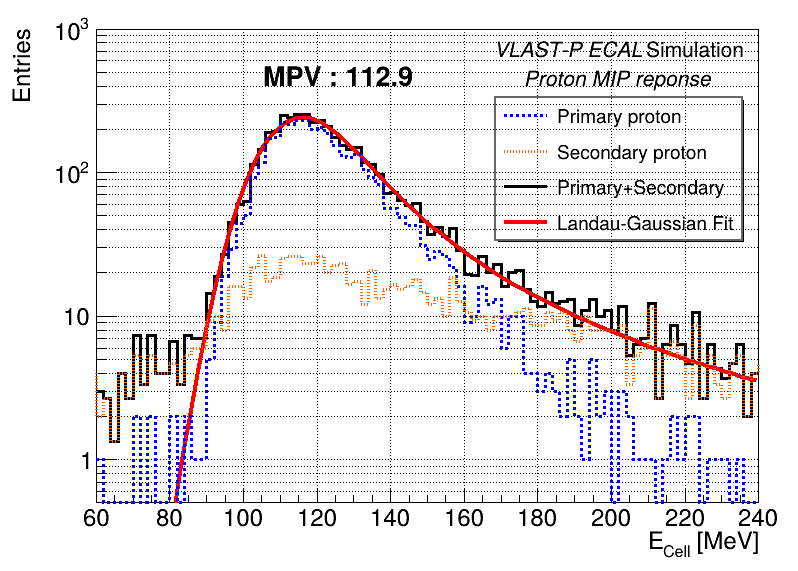}
        \caption{Proton MIP energy response in a single sensitive unit of the VLAST-P ECAL}
        \label{fig:Cell_12_Proton}
    \end{subfigure}
    \hspace{0.02\textwidth} 
    \begin{subfigure}{0.48\textwidth}
        \centering
        \includegraphics[width=\linewidth, keepaspectratio]{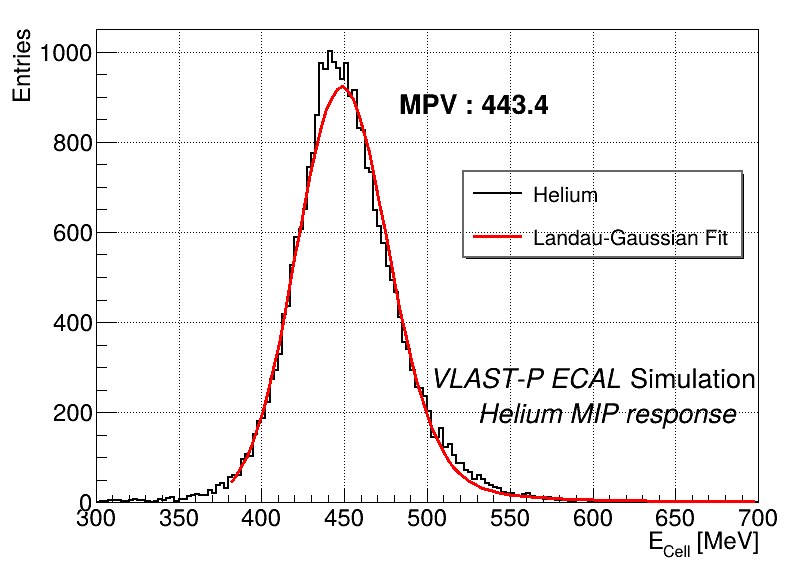}
        \caption{Helium nuclei MIP energy response in a single sensitive unit of the VLAST-P ECAL}
        \label{fig:Cell_12_Helium}
    \end{subfigure}

    \caption{MIP energy response in a single sensitive unit of the VLAST-P ECAL}
    \label{fig:Cell_12}
\end{figure}

In the actual space environment, secondary particles, most of which are protons, also exist under space conditions, typically with energies below $10\,\mathrm{GeV}$. These secondary particles originate from interactions between high-energy primary cosmic rays and atmospheric atoms, producing cascades of new particles through nuclear reactions. According to the AMS proton flux measurements, the integral flux of secondary protons accounts for approximately 40\% of the total integral flux of all protons in the equatorial region~\cite{SecondaryProton}. In this study, secondary protons are modeled using the kinetic energy distribution measured by AMS, with arrival directions sampled isotropically from a source sphere. The simulation results show that the hit multiplicity distribution follows an approximately exponential decay, indicating that most secondary protons deposit energy in only a few crystals. The energy distribution of secondary protons is broader compared to that of primary protons. The selection efficiency of secondary protons is about half of the primary efficiency. These secondary events are combined with the primary proton events in proportion to their respective integral fluxes.

The on-orbit MIP calibration results show that the energy deposition spectrum of protons follows a Landau-like distribution, with the most probable value (MPV) measured at $112.9\,\mathrm{MeV}$ in a single CsI(Tl) unit, as shown in figure~\ref{fig:Cell_12_Proton}. In the case of helium nuclei (He), it should be emphasized that their flux in space is also considerable, making them suitable for detector calibration. Compared with protons, helium nuclei are almost free from secondary components, as the production of secondary helium through nuclear interactions is extremely rare. A comparable distribution is obtained for helium nuclei, whose MPV is $443.4\,\mathrm{MeV}$, nearly four times that of the proton MIP, depicted in figure~\ref{fig:Cell_12_Helium}. This scaling behavior can be naturally understood from the Bethe--Bloch formalism, where the stopping power of charged particles in matter is approximately proportional to the square of their charge number ($Z^2$). As helium nuclei carry twice the charge of protons ($Z=2$), their energy deposition is therefore expected to be enhanced by a factor of \replaced{about}{} four, in good agreement with the observed ratio.\replaced{}{Furthermore, observations suggest that secondary cosmic rays mainly come from regions near the sources and from the general interstellar medium~\cite{SecondaryGeneral}.}

\section{Conclusion And Outlook}
\label{sec:conclusion}
In this work, we present and validate a comprehensive on-orbit calibration methodology for the electromagnetic calorimeter of the VLAST-P satellite, whose primary objective is to reproduce realistic MIP calibration under space conditions. Utilizing detailed Geant4-based Monte Carlo simulations, we systematically investigate the interactions of the main components of the spatial cosmic ray, namely protons and helium nuclei, with the detector system.

To perform on-orbit calibration, we used a database based on geomagnetic backtracing to determine the cutoff rigidity required for charged particles to reach the satellite. Using this information, we created a realistic environment for simulating minimum ionizing protons, which are used to calibrate each ECAL channel. We also applied a set of selection criteria to ensure that the events used for calibration are clean and reliable. These criteria include trigger logic selection, hit number cuts, and path length corrections. The results show that, following the application of our calibration procedure, the proton and helium MIP spectra can be accurately calibrated with a spherical incident particle source, with an efficiency of approximately 2.95\%. By additionally applying the path length correction, the MIP spectra can be accurately adjusted to their original appearances. This approach ultimately produces well-calibrated MIP spectra for all channels, confirming the effectiveness of the method.

\replaced{To quantify the on-orbit calibration time requirement, the incident proton flux was taken as \(\Phi_p = 0.015\ \mathrm{cm^{-2}\ s^{-1}\ sr^{-1}}\), 
obtained by approximately integrating the AMS-measured differential proton flux in the equatorial region~\cite{SecondaryProton}. This value is representative of near-equatorial geomagnetic cutoff conditions.}{To quantify the on-orbit calibration time requirement, the incident proton flux was taken as 
\(\Phi_p = 0.015\ \mathrm{cm^{-2}\ s^{-1}\ sr^{-1}}\), representative of the near-equatorial geomagnetic cutoff conditions. }
For each orbital revolution, an effective observation window of approximately 
\(t_{\mathrm{eff}} = 1200\ \mathrm{s}\) 
is available within the target geomagnetic latitude range (\(-20^\circ < \mathrm{lat} < 20^\circ\))~\cite{DAMPE20}, 
based on an orbital period of \(T_{\mathrm{orbit}} = 1.5\ \mathrm{h} = 5400\ \mathrm{s}\). 
Based on Monte Carlo simulations employing a spherical source with a radius of $0.5\,\mathrm{m}$, the geometric factor of the electromagnetic calorimeter for MIP calibration is calculated to be $G = 6.34 \times 10^{-3}~\mathrm{m^2 \cdot sr}$. Taking into account the proton flux and the orbital period mentioned above, this results in an expected yield of approximately 1142 valid calibration events per orbit. Considering the combined flux of primary and secondary protons, a sample of approximately $3\times10^{3}$ events per channel is required across the 25 readout channels to constrain the statistical uncertainty to $0.5\%$. In total, approximately $7.5 \times 10^{4}$ events are required. Consequently, the total acquisition requirement amounts to approximately 65.7 orbital periods, corresponding to about 98 hours (\(\approx 4.1\) days) of effective data-taking. 
\replaced{This estimate serves as a practical baseline for scheduling and planning the calibration campaigns during the operational phase of the mission.The dominant sources of systematic uncertainty are expected to arise from the on-orbit operational environment. For example, temperature variations may affect the detector response; according to communication with the satellite system team, the temperature variation within one orbit is controlled within 0.5 degree, which may introduce an uncertainty of about $1\%$ in the calibration precision. In addition, uncertainties in the geomagnetic field model and the effect of solar modulation can also contribute to systematic deviations. Based on the experience of the DAMPE Collaboration, these effects may introduce an additional uncertainty at the level of approximately $1\%$~\cite{System_error}.}{}

In conclusion, we present an effective on-orbit calibration strategy. A key feature of our work is using a geomagnetic backtracing database instead of a standard random spherical source simulation. This makes the simulation more realistic by including accurate geomagnetic cutoffs for the incident particles. MIP calibration is critical for the accurate energy reconstruction of gamma rays. Moreover, this study establishes a solid framework for future applications in the full-scale VLAST mission.

\section*{Acknowledgments}
Special thanks to the Space Exploration Program under Grant No. GJ11050103, GJ11050108, and the National Natural Science Foundation of China under Grant No. 12227805, 12273120.

\bibliographystyle{JHEP}
\bibliography{references}

\end{document}